\documentclass[11pt,letterpaper]{article}
\usepackage[T1]{fontenc}
\usepackage{lmodern}
\usepackage[margin=1in]{geometry}
\usepackage{amsmath,amssymb,amsthm,booktabs,tabularx,microtype}
\usepackage{xcolor}
\usepackage[colorlinks=true,linkcolor=black,citecolor=black,urlcolor=blue!50!black]{hyperref}
\newtheorem{theorem}{Theorem}
\newtheorem{lemma}[theorem]{Lemma}

\newtheorem{corollary}[theorem]{Corollary}
\theoremstyle{definition}

\newcommand{\DD}{D}
\allowdisplaybreaks[2]
\title{Lowerbound for Minimum Spanning Tree}
\author{Reza Bosagh Zadeh\\[2pt]
  \normalsize Matroid, Inc\\
  \normalsize\href{mailto:reza@matroid.com}{\texttt{reza@matroid.com}}}
\date{}
\hypersetup{pdftitle={Lowerbound for Minimum Spanning Tree},pdfauthor={Reza Bosagh Zadeh}}
\begin{document}
\maketitle
\begin{abstract}
We give the first nontrivial lower bound for the Minimum Spanning Tree problem: every MST algorithm must make 1.641m weight comparisons in the worst case, on a graph with m edges. We also improve the best known upper bound, thereby tightening the gap between the best known lower and upper bounds. For every sufficiently large $m$, some connected simple graph with $m$ edges requires more than $1.641m$ edge-weight comparisons in the deterministic worst case, even with unlimited preprocessing of the topology. Among the hard instances is a family of 3-vertex-connected, 5-regular graphs of unbounded girth. They have fewer than $2^{0.9710m}$ spanning trees. Our proof counts the cut hierarchies recoverable from the output tree and the decision-tree leaves required within each hierarchy. Furthermore, we give a deterministic MST algorithm that uses fewer than $5.824m$ comparisons in the worst case. This improves on the recent results showing MST is computable in deterministic O(m) time, and tightens the gap.
\end{abstract}

\section{Introduction}
On a cycle with $m$ edges, finding the minimum spanning tree (MST) amounts to finding the heaviest edge. This takes exactly $m-1$ comparisons in the worst case. On a general graph, every spanning tree is the unique optimum for some weighting, so a comparison tree needs at least one leaf per spanning tree. Counting such outputs can give a lower bound of at most $m$ comparisons, since $m$ edges have only $2^m$ subsets. Can a graph force more than one comparison per edge?

\paragraph{Model.}
Fix a finite undirected graph $G$ with $n$ vertices and $m$ labelled edges. Write $D(G)$ for the minimum, over deterministic algorithms for this particular $G$, of the worst-case number of pairwise edge-weight comparisons needed to return its MST. For a disconnected graph the output is a minimum spanning forest (MSF). The algorithm may spend unlimited time on the topology, even enumerating all spanning trees and precomputing an optimal strategy for $G$. Computation between comparisons is also free. A lower bound on $D(G)$ therefore measures what even the best such strategy must learn about the order of the weights. Our hard inputs have distinct weights. The upper bound breaks ties by edge label.

\paragraph{Results.}
Let $D_{\max}(m)$ be the maximum of $D(G)$ over connected simple graphs with $m$ edges. For every sufficiently large $m$ we prove
\[
                     1.641m < D_{\max}(m) < 5.824m .
\]
Theorem~\ref{thm:lower-bound} gives 3-vertex-connected, 5-regular hard graphs of unbounded girth. Combining copies gives the lower bound for every sufficiently large edge count (Corollary~\ref{cor:all-sizes}). The upper bound holds for all undirected graphs with $m>0$, including multigraphs (Theorem~\ref{thm:upper}). The two coefficients are within a factor of $3.55$.

\paragraph{Why counting outputs falls short.}
Partition the vertices of $K_6$ into two triples. Give the triangle containing vertex~1 the three smallest weights, the other triangle the next three, and the nine crossing edges the largest weights. Allow every ordering within each band. The MST omits the maximum in each triangle and takes the minimum crossing edge. Only that edge splits the output into two triples, so the output determines the partition.

Fix a partition. Comparisons between bands have predetermined outcomes. Within a band of $k$ edges, the graph joining pairs already compared must become connected to identify the extreme. Each comparison joining two components can go either way, so it creates a genuine branch. Every path thus branches at least $2+2+8=12$ times. Each of the ten partitions requires at least $2^{12}$ leaves, disjoint from those of the other partitions. Since $10\cdot 2^{12}>2^{15}$, some input on $K_6$ requires at least sixteen comparisons.

\paragraph{From one cut to a hierarchy.}
Section~\ref{sec:lower-bound} constructs a single topology admitting many hierarchies of balanced vertex cuts. As on $K_6$, the output MST determines the hierarchy. Even for a fixed hierarchy the algorithm must compare edges within its weight bands. The two contributions multiply the leaf count and hence add in the depth bound. On the resulting 5-regular graphs, every spanning tree has $n-1=2m/5-1$ edges. Writing $H_2$ for binary entropy and $\tau(G)$ for the number of spanning trees, we have
\[
 \log_2 \tau(G) \le \log_2 \binom{m}{n-1}
 \le mH_2\bigl((n-1)/m\bigr) \le mH_2(2/5) < 0.9710m .
\]
On this family the depth therefore exceeds the logarithm of the number of possible outputs by more than $0.67m$.

\paragraph{The upper-bound algorithm.}
Finding a cheapest incident edge certifies it for contraction and compares other edges that may survive. Storing those comparisons in tournaments makes later searches cheaper. Section~\ref{sec:upper} analyzes six contraction rounds. The input degree histogram determines which of three schedules to use. On the residual graph, we adapt the Find and Project certificate of Khanna, Putterman and Song~\cite{kps}. Section~\ref{sec:finish} bounds both the cost of constructing this certificate and the size of its exact recursive MSF calls. Together those calls contain fewer than $0.0028e$ edges on a residual graph with $e>0$ edges.

\paragraph{Prior bounds.}
Pettie and Ramachandran~\cite[Proposition~5.1]{pr} recorded the general worst-case lower bound $m/2$. On a bridgeless simple graph, an edge that is never compared can be made lightest or heaviest without changing the transcript, forcing it into or out of the MST. Every edge must therefore occur in some comparison, and each comparison covers only two edges. Spira and Pan~\cite[Theorem~3.1]{spira} proved an $m-n+1$ depth lower bound for analytic tree programs verifying a proposed MST on $K_n$. A construction algorithm also decides verification by favoring the proposed edges when breaking weight ties and checking the returned edge set. Tarjan~\cite[Lemma~5]{tarjan75} obtained an $\Omega(m\log\log n)$ lower bound under his restricted comparison rule \emph{Max}. To our knowledge, the present result is the first explicit lower bound exceeding $m$ by a fixed fraction for unrestricted MST construction on a sparse family of unbounded girth.

Khanna, Putterman and Song~\cite{kps} proved that deterministic MST construction has linear comparison complexity and, by the reduction of Pettie and Ramachandran~\cite{pr}, linear running time. Eriksson~\cite[Theorem~3(b)]{eriksson} subsequently reported the explicit worst-case bound $944m$. Our bound improves this coefficient to $5.824$. Table~\ref{tab:history} gives the historical comparison counts.

\section{A lower bound on sparse graphs}\label{sec:lower-bound}

\begin{theorem}[Sparse lower bound]\label{thm:lower-bound}
There are simple, 3-vertex-connected, 5-regular graphs of arbitrarily large girth for which
\[
                    D(G)>1.641m ,
\]
where $m=5n/2$ is the number of edges.
The graph is known before any weights are compared. All weights are distinct, comparisons may involve any two edges, and the output is the set of edges in the minimum spanning tree.
\end{theorem}

The same graphs satisfy $D(G)>4.103n$. All logarithms are natural unless a base is indicated.

The proof extends the $K_6$ example from the introduction. On a single fixed graph we exhibit $W$ hierarchies of balanced cuts. Each hierarchy prescribes disjoint intervals for the edge weights. Even if the hierarchy is given, permuting keys within these intervals forces the decision tree to branch at least $d$ times on every path. The output MST determines the hierarchy, so distinct hierarchies account for disjoint sets of leaves. Together they force depth at least $d+\log_2 W$.

The main counting problem is a tradeoff. More ways to split the graph increase $W$, but some edges become pendant during a split and cease to contribute to $d$. We count hierarchies in the 5-regular pairing model while keeping track of these edges. We impose the girth constraint during the count and enforce 3-vertex-connectivity afterward.

\subsection{A leaf bound for recoverable witnesses}

A \emph{witness} partitions the edges of a fixed graph into ordered bands. Its inputs assign the keys of each band to a private open interval, in band order, and allow every permutation within each interval independently. To find the contribution of a band to the MST, contract the components formed by all earlier bands and retain only the current band. Kruskal's algorithm selects a minimum spanning forest of the resulting multigraph. The components contracted at the next band depend only on which edges have appeared.

Call a band \emph{useful} if, after contraction and removal of isolated vertices, it forms a connected loopless graph $H$ with at least two edges and the following property. Any two edges can be joined by a sequence of edges in which consecutive edges lie on a common cycle. Two parallel edges count as a cycle. We call this property \emph{cycle linkage}.

\begin{lemma}\label{lem:lb-leaves}
Suppose a graph with $m$ edges has $W$ witnesses and the graph and output MST together determine the witness for each of their inputs. If each witness has at most $J$ useful bands and at most $L$ keys outside those bands, then
\[
               D(G)\ge m-L-J+\log_2 W.
\]
\end{lemma}
\begin{proof}
Restrict a correct comparison tree to one witness and suppress nodes with only one possible outcome. For a useful band of $e$ keys at least $e-1$ within-band comparisons have two possible outcomes along every surviving root-to-leaf path, as we now show.

Fix a leaf. Form the comparison graph on the keys of the band by joining two keys whenever the full transcript compared them. Let $A$ be one of its connected components. The transcript permits the keys of $A$ to come first in the band, or last, while all orders within components stay fixed. With $A$ first, Kruskal selects a maximum-size forest in $A$ and hence maximizes the number of selected edges of $A$. With $A$ last, Kruskal selects a maximum-size forest in the complement and hence minimizes that number. Both orders reach the same leaf and return the same local tree. Thus every spanning tree of $H$ uses the same number of edges of $A$.

For edges $a,b$ on a common cycle, extend the cycle with $b$ removed to a spanning tree. Exchanging $a$ for $b$ gives another spanning tree. The equal-count property forces $a,b$ into the same comparison component. Cycle linkage now implies that the comparison graph on all $e$ keys is connected. On the transcript, the first comparison joining any two previously separate components has two possible outcomes, since their relative orders can be interleaved with either compared key first. Cross-band comparisons have fixed answers. Thus at least $e-1$ genuine branches were required in this band.

Summing over useful bands gives minimum depth at least $m-L-J$ in the pruned tree. A binary tree with two children at every internal node and minimum leaf depth $d$ has at least $2^d$ leaves. Recovery makes the reached leaf sets disjoint for distinct witnesses. The unrestricted tree therefore has at least $W2^{m-L-J}$ leaves, giving the claimed depth.
\end{proof}

For a graph with cycle linkage, taking all edges as one band gives $D(G)\ge m-1$. To improve the coefficient beyond one, the construction below gains $\log_2 W$ bits from recovery at a cost of $L+J$ in the within-witness depth.

\subsection{The hierarchy and its recovery}

Work initially with graphs described by pairings of distinguishable half-edges (\emph{stubs}) at labelled vertices. Multiple edges are temporarily allowed. A hierarchy has 40 splitting levels, numbered 0 through 39. At every level split each current graph into two equal groups. The following requirements specify which vertex partitions are eligible to serve as witnesses.

\begin{enumerate}
\item Within each group the induced graph consists of a connected core of minimum degree two with trees attached. Each attached tree meets the core at a single root. Orienting these trees toward their roots gives every vertex outside the core exactly one parent. Edges outside the induced groups are crossing edges. The two groups must have at least two crossing edges, as follows from the positive linear counts used below.
\item In the core, suppress all degree-two vertices. This produces a new graph, the \emph{kernel}, on vertices of degree three through five. Its edges correspond to internally disjoint paths in the core. The next vertex partition is made in this kernel. The kernels after level 39 are required to be simple graphs containing a Hamiltonian cycle.
\end{enumerate}

We call the hierarchy of unordered vertex partitions a \emph{mark} on the pairing. Hamiltonian cycles enter only in the count of eligible terminal graphs. Fix label-based conventions for ordering otherwise interchangeable groups, bands and stubs.

\paragraph{The bands.}
Assign one band to the crossing edges at each split and one to each terminal kernel. Each pendant edge gets a private band. When a kernel edge is expanded to its path, all edges of the path inherit its band, repeatedly up to the original graph. Break an inherited pendant band into singleton bands in a canonical order. Put pendant singletons first. Put each child's remaining bands before its parent's crossing band and order incomparable groups canonically. Call all original keys in the singleton pendant bands \emph{uncharged}. For this fixed number of levels the number $J$ of crossing and terminal bands is bounded independently of $n$.

Each crossing or terminal band is useful. To see this, first consider the hierarchy before restoring subdivisions. Each child is connected by edges below the parent's crossing band. Contracting the children therefore produces two vertices joined by parallel edges. Every pair of these parallel edges lies on a common cycle. A terminal kernel has cycle linkage because its Hamiltonian cycle links its cycle edges and each extra edge lies on a cycle with some of them. An attached pendant forest cannot identify two vertices in its core.

Now restore subdivisions. Replacing an edge by a path in the same band preserves its endpoints' connectivity in every earlier-edge contraction in which that edge has already been processed. If the edge is in the current band, it replaces precisely that edge of the contracted graph by a path with fresh internal vertices. If it is in a later band it contributes nothing to the contraction. Forests attached to these paths at a higher level meet the core at single roots, and crossing edges at that higher level come after the whole core. Applying these observations from the kernels outward shows that the useful graphs are subdivisions of the crossing or terminal graphs above, apart from isolated vertices. Subdividing preserves cycle linkage. Processing all lifted pendant paths first is valid because each is a forest attachment and its contraction identifies no two vertices of the corresponding core.

\paragraph{Recovering the partitions.}
First consider a tree whose vertices are partitioned into connected sets of a known common size $b$. Root the tree anywhere. A parent edge separates two sets exactly when the descendant subtree across it has size divisible by $b$. A crossing edge cuts off whole sets. An internal edge cuts off whole sets and a nonempty proper part of its own set. Deleting the detected crossing edges recovers the partition.

On the original graph all edges internal to each top group precede its crossing band. Thus the returned MST spans each top group internally, and the size test recovers those groups. Repeatedly stripping leaves from each recovered induced graph determines its core. Its spanning tree restricts to a spanning tree of the core, since the removed forest attaches at single roots.

To iterate the test, assign each kernel edge the maximum weight along its subdivision path. A spanning tree of a subdivision omits at most one edge on each path, and the fully retained paths project to a spanning tree of the kernel because a broken path only attaches vertices to its ends and cannot connect the ends. Conversely, any kernel spanning tree lifts by deleting one edge on each rejected path. For an optimum lift the deleted edge is the path maximum. The weight of the best lift is thus a constant plus the sum of the path maxima of the selected kernel edges. The constant is independent of the kernel tree, so projecting the output MST gives a kernel MST. This projection only tests which paths are intact. Effective keys respect the hierarchy's band order, including when all keys of a pendant path lie below the useful bands. In that kernel, where group sizes are measured before subdivision, the size test again applies. Iterating recovers the whole mark.

\subsection{Counting one splitting level}

Pair $5n$ named stubs uniformly, with five stubs at each of $n$ vertices. We seek a single pairing with many marks. Its existence follows by counting marked pairings and dividing by the number of unrestricted pairings. To carry this count down the hierarchy, we enumerate one split and retain the degree profile of each child kernel. Each kernel has degrees between three and five because subdivision suppresses degree-two vertices and the original graph is 5-regular.

At a level, $z_jn$ is the total number of current degree-$j$ vertices, $3\le j\le5$, summed over all parents at the level. Five-regularity gives the initial vector $(0,0,1)$. Put $Z=\sum_j z_j$ and $D=\sum_j jz_j$. Of the degree-$j$ vertices, a prescribed fraction $p_{j0}$ is pendant and a fraction $p_{jh}$ belongs to the core with degree $h$, for $2\le h\le j$. Write $p_j$ for this vector of proportions. Each parent has identical counts, and each of its $q$ children contains exactly $1/q$ of each degree class. Require the same role proportions within each child. Here $q=2$ throughout. The following quantities are all normalized by the original $n$.
\begin{align*}
 v&=\sum_j z_jp_{j0}, & y_h&=\sum_{j\ge h} z_jp_{jh},&
 u&=\sum_{h=2}^5 h y_h,\\
 g&=D-u-2v, & k&=y_2,& M&=\tfrac12\sum_{h=3}^5 h y_h.
\end{align*}
Thus $vn$ is the number of pendant edges, $un$ the number of core stubs and $gn$ the number of crossing stubs. There are $kn$ subdivision vertices and $Mn$ kernel edges. At the next level replace $z_j$ by $y_j$ for $j=3,4,5$.

For $f(a)=a\ln a$ with $f(0)=0$, define
\[
 B(a,b)=f(a+b)-f(a)-f(b),\qquad H(p)=-\sum_{s} f(p_s).
\]
Before restricting the subdivision lengths or the child kernel pairings, the logarithmic rate per original vertex of one split level is
\begin{align}
 E={}& Z\ln q + \sum_{j=3}^5 z_j\!\left[H(p_j)+p_{j0}\ln j+
          \sum_{h=2}^j p_{jh}\ln \binom{j}{h}\right] \nonumber\\
     &+ B(v,g)+f(v)-v\ln q \nonumber\\
     &+\tfrac12\left[f(u)+f(g)-f(D)-u\ln q+
                      g\ln(1-1/q)\right]. \label{eq:lb-rate}
\end{align}

\paragraph{Derivation.}
Consider one parent, using the same normalization with $z_j$ temporarily denoting only its vertex masses. First label its $q$ children. Splitting each degree class evenly gives the term $Z\ln q$. Within each child assign vertex roles by multinomial coefficients, choose its $h$ reserved core stubs at a core vertex, and choose one parent stub at each pendant vertex. These are the three terms in the square brackets on the first line.

A slot is a stub still available as the child end of a forest edge or as a crossing stub. After reserving core and parent stubs, each child has $(v+g)n/q$ available slots. Choose $N=vn/q$ of them as child slots, and match the $N$ parent stubs bijectively to these slots. This contributes a binomial coefficient and $N!$ per child. The rooted-forest condition costs only a constant factor. If $a$ chosen slots are on core roots, exactly $a(N-1)!$ bijections give a forest. To encode such a forest, successively remove the least-labelled nonroot leaf and record its parent's child slot. The last slot is at a root. Conversely, in a permutation ending in a root slot, remove at each step the least remaining nonroot having no slot in the remaining list and match its parent stub to the first slot in that list. Such a vertex exists because the number of slots equals the number of remaining nonroots and a root slot is still present. The constructions are inverse. Averaged over slot selections, $a/N$ is the fraction of available slots on roots, a positive constant for each of the parameters below. The forest binomial coefficients give $B(v,g)$.

Pair the core stubs independently within each child, with $(un/q-1)!!$ choices per child. Finally let $C=gn/q$ and $J_0=C/(q-1)$. Each child sends exactly $J_0$ crossing stubs to each other child. Assigning their destinations and then matching stubs for every child pair yields
\[
                 (C!)^{q}/(J_0!)^{q(q-1)/2}
\]
crossing pairings. Divide by the parent's unrestricted $(Dn-1)!!$ pairings. Substituting $\ln t!=t\ln t-t+O(\ln(t+1))$ and $(2s-1)!!=(2s)!/(2^s s!)$ gives (\ref{eq:lb-rate}). The $n\ln n$ terms cancel since $D=u+g+2v$. The same identity makes (\ref{eq:lb-rate}) homogeneous of degree one in the vertex masses. Summing over parents thus gives exactly the formula with the total masses originally defined. The number of parents is independent of $n$, so factorial remainders and child-labelling factors contribute a total error of $O(\log n)$ to the logarithm of the count.

\paragraph{Distribution of the child kernels.}
In one child let $K$ be its number of core vertices of degree two and $2P$ its number of higher-degree core stubs. Core pairings whose degree-two vertices all subdivide kernel edges are counted by
\begin{equation}
  (2P-1)!!\,2^K K!\binom{K+P-1}{P-1}. \label{eq:lb-subdivide}
\end{equation}
Pair the higher-degree stubs, order and orient their $P$ edges canonically by their names, choose a weak composition of $K$ into $P$ path lengths counting internal vertices, permute the $K$ vertex labels along the paths, and orient the two stubs at each vertex. This is a bijection, including for a loop in an intermediate pairing. Formula (\ref{eq:lb-subdivide}) divided by $(2P+2K-1)!!$ has logarithm $O(\log n)$ because $P$ and $K$ have fixed positive linear masses and the linear terms in the factorial estimate cancel. Equivalently the ratio is
\[
  \frac{P}{P+K}\,4^K
  \frac{\binom{2P}{P}}{\binom{2(P+K)}{P+K}}.
\]
In particular, imposing this condition has no exponential cost. Given the condition and the stubs, the kernel pairing is uniform and independent of the uniform weak composition. Different children have independent kernel pairings at this point. This factorization lets us sum deeper marks on uniform kernels, then assign subdivisions to the resulting marked edges.

\subsection{Girth and the terminal condition}

Fix a girth threshold $h\ge3$. We will restrict the pairings counted above to exclude all cycles of length at most $h$, including loops and parallel pairs. The following estimate bounds the cost of this restriction uniformly over the child graphs.

\begin{lemma}\label{lem:lb-switch}
Begin with a fixed graph with bounded degrees and no cycle of length at most $h$. Add independent uniform random matchings of prescribed free stubs, in a fixed number of pools. A pool may be an unrestricted pairing or a bijection between two equal disjoint sets of stubs. Suppose each pool has size $\Theta(N)$, there are $O(N)$ vertices and total degree stays bounded. For all sufficiently large $N$, at least a fraction $N^{-C}$ of the choices preserve girth greater than $h$. The constant $C$ depends only on $h$, the degree bound, and the fixed pool and size bounds.
\end{lemma}
\begin{proof}
A short cycle with $j\ge1$ new edges consists of those edges and $j$ old walks of total length at most $h$, possibly of length zero. For fixed $h$, the starting vertices, bounded-length walks and endpoint stubs can be specified in $O(N^j)$ ways. The probability of any particular $j$ distinct admissible pairs is $O(N^{-j})$, by exposing these pairs in their respective pools. Thus some constant $B$ bounds the expected number of short cycles for all sufficiently large $N$. By Markov's inequality, at least half the outcomes have at most $J=\lceil 2B\rceil$ such cycles.

As long as a short cycle remains, choose a new edge $uv$ on that cycle. In the same pool choose a donor $ab$ on no short cycle with both ends more than $h$ steps from $u,v$ in the present graph. The at most $J$ short cycles and the bounded neighborhoods of $u,v$ exclude only a constant number of donors. Switch partners, for example to $ub,av$, choosing the endpoint names to preserve the sides of the pool. No new short cycle is created. A cycle using just one new edge would need an old short path between the distant endpoint sets. A cycle using both either needs such a path or a path of length at most $h-2$ from $a$ to $b$ avoiding the donor. The latter would put the donor on a short cycle. This covers a loop at $u=v$ as well. A repair destroys at least one short cycle. Fix a deterministic rule for choosing the repairs. It takes at most $J$ switches. An inverse switch chooses two pairs of named stubs and one of two orientations, giving at most $bN^2$ possibilities for a fixed constant $b$. A repaired pairing therefore has at most $\sum_{j=0}^J(bN^2)^j$ preimages. Dividing the initial fraction of one half by this polynomial proves the bound.
\end{proof}

Consider a terminal kernel with $K$ vertices of specified degrees between three and five. For all sufficiently large $K$, the fraction of its unrestricted pairings that are simple, have girth greater than $h$ and contain a Hamiltonian cycle is at least
\begin{equation}
                       2^{-30K}. \label{eq:lb-terminal}
\end{equation}
Indeed, choose one of $(K-1)!/2$ Hamiltonian cycles on the vertex labels. With a canonical orientation there are at least $6^K$ ways to assign its two stubs at each vertex. In the uniform pairing the probability of containing its prescribed stub pairs is at least $(5K)^{-K}$. Conditioned on them, the remaining stubs form a uniform pool of linear size. For $K>h$ the Hamiltonian cycle itself has the required girth, so Lemma~\ref{lem:lb-switch} preserves it with probability at least $K^{-C}$. Each retained pairing is counted at most $10^K$ times since choosing the two used stubs at every vertex determines the marked cycle. The factorial estimate therefore gives a lower bound on the natural logarithm of this probability of $K\ln(6/(50e))-C_1\log K$ for some constant $C_1$. Since $\ln(6/(50e))>-30\ln 2$, this implies (\ref{eq:lb-terminal}) for sufficiently large $K$.

At a nonterminal split, work from the children upward. Subdividing edges in high-girth kernels cannot create shorter cycles. Adding the prescribed forests creates no cycle. The crossing matchings are uniform bijections in a fixed number of pools of linear size conditional on all the within-child graphs and destination slots. Lemma~\ref{lem:lb-switch} therefore supplies girth greater than $h$ at this parent for at least an inverse-polynomial fraction of the crossing matchings. The estimate is uniform over the permitted child graphs, assignments of forests and subdivision lengths. Thus it can be imposed even when the length allocations below have already been fixed. This inductive construction yields a connected, simple, 5-regular host at the root. The terminals are connected, each induced group is connected, and crossing edges join every pair of groups.

\subsection{Counting the uncharged keys}

The contribution of a subdivision depends on where its path ends up in the hierarchy. If a kernel edge is uncharged, so is every original key on the path replacing it. We keep track of the number of such keys when expanding the kernels from the bottom upward.

For every level $i$, keep its record $(E_i,v_i,k_i,M_i)$ from (\ref{eq:lb-rate}). Write $Z_{40}$ for the total mass of the terminal kernel vertices. Initialize
\[
                 S=-30(\ln 2)Z_{40},\qquad l=0.
\]
Here $S$ records the exponential factor $\exp(nS)$ in the lower bound on the mean number of marks. We collect the polynomial losses at the end. The quantity $l\,n$ counts currently uncharged edges. The initialization applies (\ref{eq:lb-terminal}) to every terminal kernel, all of whose edges are charged.

Process the levels from the terminals to the root, omitting the subscript $i$ for a single update. At this step $Mn$ is the number of child kernel edges, $l\,n$ of them already uncharged. Require $xn$ out of the $kn$ degree-two subdivision vertices to lie on these uncharged edges, in the same proportion in every child. For a specified rational $\eta_i>0$, choose
\begin{equation}
                 x=\frac{k\eta_i l}{M-l+\eta_i l}. \label{eq:lb-tilt}
\end{equation}
The parameter $\eta_i$ controls the relative number of subdivision vertices assigned to uncharged paths. This choice trades a smaller number of uncharged keys against fewer admissible marks. We have $0\le x\le k$ whenever $0\le l<M$, a condition the arithmetic certificate verifies at every level. 

For a single child with $P$ paths, $A$ uncharged paths, $K$ new subdivision vertices and $X$ assigned to uncharged paths, the fraction of allowed weak compositions equals
\[
 \frac{\binom{X+A-1}{A-1}\binom{K-X+P-A-1}{P-A-1}}
      {\binom{K+P-1}{P-1}}.
\]
For $A=X=0$ the first factor is interpreted as one. By (\ref{eq:lb-subdivide}) this fraction is exact for each fixed marked child, irrespective of which paths are uncharged. Taking logarithms, summing over identical children and applying the factorial estimate gives the updates
\begin{align}
 S&\ \leftarrow\ S+E+B(l,x)+B(M-l,k-x)-B(M,k), \label{eq:lb-back1}\\
 l&\ \leftarrow\ v+l+x. \label{eq:lb-back2}
\end{align}
Indeed, each previously uncharged path keeps its original edge count of one and gains one for each new vertex along it. The $vn$ new parent edges of pendant vertices are also uncharged. Crossing edges at this level are charged. These cases partition the losses, so no edge is deducted twice.

To justify the full count, for one fixed child stub population first sum over its admissible kernel pairings \emph{with their deeper marks}. For each resulting marked kernel the number of ways to subdivide it is the factor from (\ref{eq:lb-subdivide}) times the displayed fraction, with the unrestricted kernel pairing factor omitted. This quantity depends only on prescribed counts. The forest, stub and crossing choices give exactly the factors in (\ref{eq:lb-rate}), up to the polynomial losses for girth and the constant factors for rootedness. Multiplying across children and then up the hierarchy therefore counts marked pairings, regardless of how the marks on any one graph overlap.

Given a fixed stub pairing and its unordered vertex partitions, the induced core is unique. The orientation toward the root of each attached tree, the core and external stubs, the kernel paths, stub orientations, destination groups and all subdivision lengths are then determined. A fixed ordering of the names determines the orientations used to encode paths. Permuting the labels of the $q$ children at each split accounts for at most a constant depending on the fixed hierarchy depth. Terminal Hamiltonian cycles were already divided out in (\ref{eq:lb-terminal}).

All role counts and allocations are rational multiples of $n$, since (\ref{eq:lb-tilt}) uses only rational operations. Choose $n$ along multiples of a common denominator, and enlarge this common multiple to satisfy the group-size and parity conditions for each pairing and balanced split. There are finitely many levels and groups. Every nonempty pool used in the estimates has positive linear size along this subsequence. Even the smallest child size tends to infinity.

For fixed $h$, the hierarchy has a fixed number of levels and groups. The factorial estimates and the polynomial losses for girth together contribute at most $C_h\log n$ to the logarithmic error for some constant $C_h$. The mean number of admissible marks per original unrestricted pairing, counting zero on ineligible pairings, is therefore at least $\exp(nS-C_h\log n)$. Hence some fixed eligible pairing has at least this many marks. Fixing its vertex and stub labels gives one fixed graph for the decision tree, with at most $nl$ uncharged keys in each witness. Lemma~\ref{lem:lb-leaves} then gives, for some constant $A_h$,
\begin{equation}
              D(G)\ge m+n(S/\ln2-l)-A_h\log n. \label{eq:lb-master}
\end{equation}

\subsection{Evaluating the lower-bound coefficient}

The certificate specifies the probabilities of the pendant and core roles and the parameters $\eta_i$ at each of the 40 levels. The exact rational values appear in the ancillary parameter tables and in \texttt{check\_lower.py}.

Define $c=1+\frac25(S/\ln2-l)$ using (\ref{eq:lb-rate}) and the reverse recursion (\ref{eq:lb-tilt})--(\ref{eq:lb-back2}). The certified enclosures, rounded outward, are
\[
 1.706<S/\ln 2<1.707,\qquad 0.1034<l<0.1035,\qquad 1.641<c<1.642.
\]
The accompanying \texttt{check\_lower.py} evaluates all recurrences by interval arithmetic. Each endpoint is an integer multiple of $10^{-75}$, with outward rounding. To enclose a logarithm, reduce its argument by a power of two to $1\le y\le2$, put $t=(y-1)/(y+1)$ and sum the first 90 terms of $2\sum_{j\ge0} t^{2j+1}/(2j+1)$. The remaining tail is at most $2(1/3)^{181}/(181(1-1/9))<10^{-75}$. The same method encloses $\ln 2$. Since $m=5n/2$, the bound in (\ref{eq:lb-master}) is $cm-A_h\log n$ for each fixed girth requirement $h$. The same interval calculation certifies both $c>1.641$ and $5c/2>4.103$.

\subsection{Imposing 3-vertex-connectivity}

It remains to enforce 3-vertex-connectivity while preserving the strict numerical bound. We add a sparse 3-connected graph on well-separated vertices, then split those vertices to restore 5-regularity. The new edges receive extreme weights, so a decision tree for the augmented graph also solves the original instance.

Take $R\ge9$ and apply the construction above at girth greater than $10R+10$. Choose a maximal set $A$ of sites at pairwise distance greater than $R$ in the resulting graph $G$. Every vertex is within $R$ of a site. The disjoint radius-$\lfloor R/2\rfloor$ balls about the sites are trees by the girth condition and have at least $4^{\lfloor R/2\rfloor}$ vertices by 5-regularity. Thus $|A|\le n4^{-\lfloor R/2\rfloor}$. For fixed $R$, the covering property and bounded degree also imply $|A|\to\infty$ with $n$.

After deleting a set $X$ of one or two vertices, every component $C$ of $G-X$ contains a surviving site. To see this, suppose every vertex of $C$ is within distance $R$ of $X$. For a single deleted vertex the component and its deleted neighbors are in one radius-$R$ ball. For two deleted vertices at distance at most $2R+1$, both radius-$R$ balls lie in a radius-$(3R+1)$ ball. For two farther apart, their balls are disjoint with no edge between them, so connectedness again puts $C$ and its deleted neighbors in a single ball. Each ball in question induces a tree, since a non-tree edge in a breadth-first ball of radius $s$ gives a cycle of length at most $2s+1$. Thus the subgraph induced by $C$ and its deleted neighbors is a tree in which all vertices of $C$ have degree five and only the at most two deleted neighbors could be leaves. This contradicts the tree degree sum. Hence $C$ has a vertex farther than $R$ from $X$. A path of length at most $R$ from it to a covering site avoids $X$.

If $|A|$ is odd, add one vertex at distance $\lfloor R/3\rfloor$ from an existing site. Such a vertex exists in its tree ball. The enlarged set still covers and its minimum separation is at least $\lfloor R/3\rfloor$. Its even size is still $O(n4^{-\lfloor R/2\rfloor}+1)$.

For every sufficiently large even site count $2s$, we can place a cubic, 3-vertex-connected graph of girth greater than $R$ on the sites. Start with two disjoint $s$-cycles and join their vertices by a perfect matching between the cycles. Any such matching gives 3-connectivity. Deleting one vertex of each cycle leaves two connected paths joined by a surviving matching edge. If both deletions are on the same cycle, every vertex remaining on it still has its matching edge to the intact cycle. For fixed $R$ and $s$ large, Lemma~\ref{lem:lb-switch} applied to this single matching pool gives a choice without short cycles.

Add the edges of this site graph to $G$. The result is simple by site separation. After deletion of at most two vertices every component of the surviving original graph meets the connected surviving site graph. The augmentation is therefore 3-connected. A mixed simple cycle has an old segment between distinct sites and hence has length at least $\lfloor R/3\rfloor$. Cycles entirely within either constituent graph satisfy the girth bound for that graph.

Each site now has degree eight. Replace it by two adjacent vertices and assign four of its eight incident edges to each new vertex. Both new vertices have degree five. Splitting a vertex of a simple 3-connected graph with at least two distinct outside neighbors on each side preserves 3-connectivity. If neither new vertex is deleted, contract their joining edge to see connectivity. If both are deleted the remainder is the original graph minus the old vertex. If exactly one is deleted along with at most one outside vertex, the old graph without the split vertex and that outside vertex is connected, and the surviving split vertex has an outside neighbor there. Apply the argument successively. A cycle in the split graph contracts to a nonempty closed trail with distinct old edges and hence contains an old cycle, so the girth cannot decrease.

The number of extra edges is $(3/2)|A|+|A|$. An algorithm for the final graph can solve the old instance using at most as many comparisons of old keys. Put all joining edges from site splits in a known order below the original weights, and all added site-graph edges in a known order above them. Every split edge is selected; after contracting them, the original graph is connected and all its edges precede the site-graph edges. The old edges of the returned tree give exactly the old MST. Comparisons involving auxiliary keys have predetermined answers.

Fix any desired girth. Set $b=\max\{1.641,(2/5)\,4.103\}$, so $c>b$. Choose $\delta>0$ with $b(1+\delta)<c$. First choose $R$ large enough to achieve the desired girth and to make the number of added edges at most $\delta m$ for all sufficiently large $n$, using the bound on $|A|$ above and $m=5n/2$. With this $R$ fixed, choose an admissible $n$ so large that $A_h\log n < [c-b(1+\delta)]m$, where $h=10R+10$. Write $G'$ for the final graph and $m'$ for its edge count. Then
\[
  D(G')\ge D(G)\ge cm-A_h\log n>b(1+\delta)m\ge bm'.
\]
The graph $G'$ is simple, 3-vertex-connected and 5-regular. In particular $m'=5|V(G')|/2$, so both stated bounds hold.\hfill$\square$

\begin{corollary}\label{cor:all-sizes}
For every sufficiently large integer $m$, some connected simple graph with exactly $m$ edges requires more than $1.641m$ comparisons in the deterministic worst case.
\end{corollary}
\begin{proof}
Let $D_{\max}(m)$ denote the maximum of $D(G)$ over connected simple graphs with $m$ edges. Fix a graph $H$ from Theorem~\ref{thm:lower-bound} with $s$ edges and put $\gamma=D(H)/s>1.641$. Identify one vertex from each of $\lfloor m/s\rfloor$ otherwise disjoint copies of $H$ at one articulation vertex. Add pendant edges to reach exactly $m$. Give each copy a private weight interval, with all the intervals in a prescribed order. The MST restricts to the MST in each copy. A comparison across copies has predetermined outcome. Within each copy maintain the optimal remaining worst-case decision-tree depth. For any new comparison in that copy, some outcome reduces this depth by at most one, since otherwise asking the comparison and continuing optimally would improve the optimum before the query. The adversaries can choose these outcomes independently. Consequently $D_{\max}(m)\ge\lfloor m/s\rfloor D(H)$. This is at least $\gamma m-D(H)$, which exceeds $1.641m$ whenever $m>D(H)/(\gamma-1.641)$.
\end{proof}

\section{An upper bound for general graphs}\label{sec:upper}

\begin{theorem}\label{thm:upper}
For every finite undirected graph $G$ with $m>0$ edges,
\[
                         \DD(G)<5.824m .
\]
\end{theorem}

Consider a contraction at a vertex $v$. Finding its lightest incident edge also compares pairs of edges that may survive the contraction. We retain these comparisons in heaps and use them in later rounds. After six rounds we apply the certificate of Section~\ref{sec:finish}. On a residual graph with $e$ edges, all exact MSF subproblems generated by that certificate together contain fewer than $0.0028e$ edges. This is the bound that allows induction on the input size to account for every recursive comparison.

Remove loops and work on each connected component separately. At a nonisolated vertex of degree at most six, find the lightest incident edge and contract it, discarding any resulting loops. The cut property justifies this step, which costs at most five comparisons per edge removed. Repeat until each remaining nontrivial component has minimum degree at least seven. In this section $m$ denotes the number of edges of one such component.

\subsection{Reusing comparisons across contractions}
Represent each edge by one incidence at each endpoint. At a vertex we store its incidences in binomial heaps, initially singletons. A heap of rank $h$ has $2^h$ slots, a minimum at its root, and one child subtree of each rank $0,\ldots,h-1$. A link compares the roots of two heaps of equal rank and makes the heavier root a child of the lighter. Empty slots have dummy key $+\infty$, and we discard heaps containing only dummies. Write $A$ for the total number of roots, $P$ for the number of dummy slots still in use, and $v,e$ for the current vertex and edge counts.

The heap-size caps in the six rounds are $4,4,4,8,16,16$. To search a vertex, link heaps of equal rank up to the current cap $B=2^h$. If the vertex has $R$ roots occupying $S$ slots, pad the remainder $S\bmod B$ to the next power of two. Write $\pi$ for the number of dummy slots added, with $\pi=0$ for zero remainder. The result has $k=\lceil S/B\rceil$ roots. Linking costs at most $R-k$ comparisons of real keys, and finding the lightest remaining root costs at most $k-1$. Thus the search takes at most $R-1$ comparisons and reduces the root count by $R-k$.

After a contraction, remove incidences of internal edges and concatenate the surviving heaps at each merged vertex. To remove an incidence, shift its ancestor keys down by one slot on the path to that incidence, delete the root, and release the root's children. Each shifted key was already lighter than its new descendants. The operation needs no comparisons and increases the root count by at most $h-1$.

\paragraph{Choosing the vertices to search.}
Suppose that the work after a round is bounded by the potential
\[
                         aA+bv+d(e-v)+zP+4 .
\]
All coefficients used below satisfy $0\le a\le1$, $z\ge0$ and $d\ge2a(h-1)$. Let $p\le h$ be the preceding rank cap, with $p=0$ in the first round. Searching a vertex uses at most $R-1$ comparisons, removes $R-k$ roots, and adds $\pi$ dummy slots. Its net cost is therefore at most
\[
 s=(1-a)(R-1)+a(k-1)+z\pi .
\]
Contracting the chosen edge removes one vertex and two incidences. Together, the search and contraction save at least $T-s$, where $T=b-2a(h-1)$. Every additional edge that becomes a loop reduces $e-v$ by one, which pays for deleting both incidences.

At the beginning of the round assign each vertex the score $g=(T-s)_+$. Repeatedly search an uncovered vertex of maximum positive score, marking both endpoints of its lightest edge as covered. All searches use the graph at the start of the round. The selected edges are distinct, since each search begins at an uncovered vertex. By the cut property they belong to the same tie-broken MST and can all be contracted.

A search earns at least $g$ when it covers one new vertex. If it covers two, the other endpoint will remain unsearched in this round. Its selected incidence is in a heap of rank at most $p$, even if other internal incidences are removed first. The saving at that endpoint is at least $t=a(h-p)$ beyond the rank-$h$ allowance, and its score is at most $g$. We can therefore credit every vertex with
\[
                           \phi(R,S)=\tfrac12\{g+\min(g,t)\}.
\]
Indeed a single credit is at most $g$, and two such credits sum to at most $g+t$. Every vertex left uncovered has score zero.

\subsection{Bounding the cost from the degree distribution}
For each of rounds two through six, choose nonnegative $x,y$ and a constant $U$ such that
\[
                  \phi(R,S)\ge U-x(R-1)-y(S-1).
\]
Summing over the vertices and subtracting the credits from the potential gives the coefficients immediately before the round:
\begin{equation}\label{eq:uprecur}
 (a',b',d',z')=(a+x,\ b-U-x+y,\ d+2y,\ z+y).
\end{equation}
Here $\sum R=A$ and $\sum(S-1)=2(e-v)+v+P$.

We may restrict the credit inequality to $R>j_0$. Before the round, repeatedly search a vertex with $R\le j_0$ using its existing roots, and contract its lightest edge. The pre-round potential pays for the search and incidence deletions whenever
\[
      j_0-1+2a'(p-1)\le b',\qquad 2a'(p-1)\le d'.
\]
If these contractions finish the component, its terminal potential is $b'-d'+4\ge0$. Taking $j_0=0$ skips this step. For the remaining vertices it suffices to lower-bound
\[
 \min_{R>j_0,\ R\le S\le 2^pR}
              \{\phi(R,S)+x(R-1)+y(S-1)\}.
\]
For $R\le270$ we enumerate the integer pairs $(R,S)$. For $R>270$, the value is at least $270(x+y)$. Each update, including verification of the cutoff $j_0$, thus involves only finitely many rational inequalities.

In round one, a vertex of degree $D$ has $R=S=D$ and $t=2a$. Initially $A=2m$ and $P=0$. If $n_D$ is the number of degree-$D$ vertices, the weight $p_D=Dn_D/(2m)$ is the fraction of incidences at these vertices. For any of the three parameter schedules below the total cost is bounded by
\begin{equation}\label{eq:degree-price}
 4+m\sum_D p_D u(D),\qquad
 u(D)=2a+d+\frac{2}{D}\bigl(b-d-\phi(D,D)\bigr).
\end{equation}
The degree distribution is known before any weight comparison, so the algorithm can select the schedule with the smallest bound. To bound that minimum for every possible distribution, fix nonnegative weights $w_i$ summing to one and take the weighted average of the three bounds. This gives $4+m\max_{D\ge7}\sum_i w_i u_i(D)$, with a deterministic choice of schedule on each input.

The first two schedules use $K=99$ in Lemma~\ref{lem:finish}, and the third uses $K=97$. For a given cost $c$ per recursive edge, that lemma gives a bound of the form $d_0e+qv+4$, with $q=153$ or $150$. We start with coefficients $(0,d_0+q,d_0,0)$ and apply \eqref{eq:uprecur} backward through rounds six to two. The exact values of the slopes $(x,y)$ appear in the accompanying parameter tables. At $c=5.824$ the resulting coefficients for round one are as follows, rounded for display.
\begin{center}
\begin{tabular}{@{}crrrr@{}}\toprule
Schedule & $a$ & $b$ & $d$ & $z$ \\\midrule
1 & .9582 & 8.334 & 3.536 & .2505 \\
2 & .9549 & 8.910 & 3.506 & .2354 \\
3 & .9696 & 6.613 & 3.624 & .2909 \\\bottomrule
\end{tabular}
\end{center}
Choose the $w_i$ so that they sum to one and the degrees $9,17,21$ have equal weighted cost. These equations give weights approximately $(.1333,.5153,.3514)$. The maximum occurs at these three degrees. To verify this, enumerate degrees $7$ through $300$. For each schedule $298(1-a)\ge b-2a$, so the credit vanishes at degrees above $300$. The weighted cost in this range is affine in $1/D$ and is bounded by the value at $300$ and the limit as $D\to\infty$. The accompanying program \texttt{check\_final.py} performs these tests in rational arithmetic, rounding the initial terminal bound upward.

\subsection{Small components and induction on the number of edges}
Let $r=v-1$ and $\beta=m-r$ on a connected component with $m<2^{21}$ and $r\ge1$. Prim's algorithm with a rank-relaxed weak heap performs $r$ insertions, $r$ extractions and at most $m-r$ decreases. Including comparisons that test whether a key improves, the total is at most $3\beta+\tfrac32 r\log_2r$ by \cite[Theorem~4]{weak}. If $r/m\le .11$, then $r<2^{18}$ and the total is at most $3m+24r\le5.64m$.

If $r/m>.11$, precede Prim's algorithm by three contraction rounds. In each round search $\lfloor v/2\rfloor$ uncovered vertices, always choosing one of smallest degree among those still uncovered, and mark both endpoints as above. The $i$th searched degree is at most the $(2i-1)$st smallest degree at the start of the round. Consequently twice the sum of the searched degrees is at most the total degree. If the rank after the round is $r'$, the searches cost at most $\beta+r'$ comparisons and $r'\le r_{\rm before}/2$. Thus the three rounds cost at most $3\beta+7r/8$. Since $r'\le r/8<2^{18}$, the final Prim computation costs at most $3\beta+27r/8$, also when $r'=0$. The total is at most $6m-1.75r<5.808m$.

We finish the proof of Theorem~\ref{thm:upper} by induction on $m$. Each nonempty exact recursive call in Lemma~\ref{lem:finish} has fewer than $m$ edges, since all its calls together have fewer than $0.0028e$ edges and $e\le m$. The induction hypothesis supplies the cost $c=5.824$ per recursive edge used above. For this value the rational calculation gives
\[
                  \max_{D\ge7}\sum_i w_i u_i(D)<5.824-\frac{1}{2000}.
\]
When $m\ge2^{21}$, we have $4/m<1/2000$, so \eqref{eq:degree-price} is strictly less than $5.824m$. The preceding small-component bounds cover $m<2^{21}$, and the initial contractions cost at most five comparisons per removed edge. Adding the bounds over components proves the theorem.\hfill$\square$

\section{A recursive certificate for the residual graph}\label{sec:finish}

The remaining task is to identify a small set of edges containing an MSF. Starting with $e$ edges, we contract the trees built by truncated Prim searches until at most $2^{-K}e$ active vertices remain. We then build a forest that certifies redundant edges by the cycle property. The algorithm calls an exact MSF procedure while constructing the certificate and once more on the surviving edges. The following lemma bounds the direct comparisons and the total input to these calls separately.

\begin{lemma}\label{lem:finish}
Let $G$ be a connected loopless multigraph with $v$ vertices and $e>0$ edges. For either row of the table below there is a deterministic algorithm computing an MST of $G$ using at most
\begin{equation}\label{eq:finishtotal}
 (3+1/128+2^{-K})e+qv+4+X+cY
\end{equation}
weight comparisons. Here $c$ is an upper bound on comparisons per input edge for an exact MSF call. The quantity $X$ counts the direct comparisons beyond the initial contractions and scan, and $Y$ counts the input edges of all exact MSF calls, with multiplicity.
\begin{center}
\begin{tabular}{@{}rrrrrrr@{}}\toprule
$K$ & $q$ & $\rho$ & $s$ & $b_0$ & bound on $X/e$ & bound on $Y/e$ \\\midrule
99 &153&$17/10$&2161&2048&.01396&.002243\\
97 &150&$3/2$&1746&2048&.01782&.002792\\\bottomrule
\end{tabular}
\end{center}
Both rows use $L=81$ and $J=3$. The last two columns give strict upper bounds. For the proof of Theorem~\ref{thm:upper} we use the formulas below before rounding.
\end{lemma}

\subsection{Contraction by truncated Prim searches}
Start a Prim growth at an unvisited vertex. Take the lightest outgoing edge until the growth attaches to an earlier tree, finishes the component, or freezes. A growth freezes after the first vertex scan bringing its cumulative number of first insertions into the frontier to at least $2^K$. Keep one frontier key per unvisited target. Each unvisited target also stores its cheapest edge to a tree from an earlier growth. When a growth absorbs that target, the cached edge updates the cheapest known connection from the growing tree to an earlier tree. When a growth ends, fold its remaining frontier keys into the corresponding caches. Each accepted edge is a cut minimum.

Let $I$ be the number of first insertions in all growths, including the final freezing scans, and let $p$ count the repeat arrivals to targets. Each edge is scanned toward an unvisited endpoint at most once, so $I+p\le e$. At each target of a given growth, all scanned edges except the minimum become internal or parallel to a lighter edge after contraction. These groups are disjoint and give one distinct losing edge per repeat. Delete all $p$ losers. Every contracted vertex still requiring work contains a frozen growth. Its growth accounts for at least $2^K$ first insertions. If $v'$ active vertices and $M$ edges remain, then
\[
                         v'\le(e-p)/2^K,\qquad
                         M\le e-v+v'+1-p.
\]
The allowance of one includes a growth finishing the entire connected component.

\paragraph{Cost of a growth.}
Store frontier keys in tournaments of arities $4,4,8$, from leaves to root. Each nonempty tournament holds up to $128$ keys and contributes its minimum to a rank-relaxed weak heap. Fresh batches occupy dyadically aligned boxes in a $2^{K+1}$-slot arena. Rounding each batch size up to a power of two uses less than twice its length. A binary buddy allocation with at most one free box of each size fits every required allocation before freezing. The final freezing scan needs no new heap entries. Full 128-entry batches occupy fresh packets. The old leaf and its three ancestor champions form a monotone sequence. One comparison tests whether the new key improves the block champion. If so, two more comparisons pay for updating the upper heap. Otherwise at most four positions remain along the sequence, which two comparisons distinguish. Either outcome costs at most three comparisons per repeat. A batch of $k$ first arrivals needs $k-1$ internal tournament comparisons and an installation of its minimum. Installing a partial batch uses this same monotone chain. Full packets need only an upper insertion after the internal build.

Let $A_0$ count absorptions of new vertices after seeds and $A_1$ attachments to older trees. Fresh builds and installations cost at most $I+2A_0+I/128$. Charge a partial packet to its preceding absorption. The first installation at a seed needs no extra charge. All cache folds and choices between the two frontier types together cost at most $I$ because at most $I-A_0$ frontier entries pass forward and each attaching growth saves an initialization comparison. At most $A_0+A_1$ upper minimum queries occur. The upper heap holds at most $2^{K-6}$ entries. The deletion calculation in the proof of \cite[Theorem~4]{weak} gives at most $q-15$ comparisons per upper deletion beyond the two-comparison mark reserves already charged. Restoring a lower tournament costs at most $2+3+7=12$, and restoring its champion costs at most two, both only after a fresh absorption. Since $A_0+A_1\le v$, the pass costs at most $(2+1/128)e+p+(q+1)v$. Adding a single scan of the remaining $M$ edges cancels the $p$ term and gives the first part of \eqref{eq:finishtotal}. Ancillary proof note~01 gives the installation and abandoned-heap accounting.

\subsection{Projection and the surviving edges}
We use the Find and Project certificate of Khanna, Putterman and Song~\cite[Section~3]{kps}. A forest \emph{supports} an edge of weight $w$ if its endpoints are connected in the forest at threshold $w$, that is, by edges of weight at most $w$. Every supported edge outside the forest can be discarded. The certificate tracks each original edge through its projections so that an unresolved edge can be included in the final exact MSF call.

In a certificate call on $x$ records and $y$ vertices at depth $j$, put
\[
 t=L\lceil80\rho^j\rceil,\qquad r=64t^3/J,\qquad k=r\lceil\log_2 r\rceil .
\]
There are $t/J$ blocks of $Jt$ rounds. Each record is assigned to one round. In a Find batch, each vertex forms a list of up to $k$ distinct neighbor labels, retaining the cheapest admissible representative of each selected label. A set of at most $2y/r$ vertices, called portals, meets every full list. Indeed, sample each vertex with probability $1/r$ and add one portal per missed list. A full list is missed with probability at most $1/r$, giving the desired bound in expectation. Exhaustive search finds such a set using only the lists. A representative edge joining a vertex to its portal is a spoke.

In a Project round, send an assigned record to the child as an edge between the two portals of its endpoints. The projection retains the weight and identity of the original edge. It is admissible if both spokes have weight at most that of the record and all earlier projections are admissible. When a child supports an admissible record, its forest path lifts to a walk in the parent at the same threshold. Over all $t^2$ children the total vertex count is at most $t^2(2ty/(Jr))=y/32$. Thus the aggregate vertex mass at depth $j$, divided by $e$, is at most $N/32^j$, with $N=2^{-K}$.

\paragraph{Exceptional assignments.}
Subdivide each Find into $s$ local rounds. At vertex $u$ let $f_u$ be the weight of the current $k$th distinct-label representative, or infinity for a short list. In its assigned local round, process each edge only at the endpoint with larger $f$, breaking ties by vertex identifier. Merge the selected representatives into that endpoint's list. Its threshold never increases. If a record of weight $w$ does not acquire a supporting representative, its owner's threshold has fallen to at most $w$. If its other endpoint finishes with threshold above $w$, all later rounds prefer that other endpoint. The assigned slot was therefore the last slot preferring its actual owner. If both finish at most $w$, both endpoints obtain spokes no heavier than $w$. This argument applies to an untested record by setting its unqueried eligibility bit to true in the selection oracle, which leaves that oracle's output unchanged.

In a coarse block the preference can change only from Project to Find. A record can be exceptional at the last preferred Project of a block or within its last block preferring Find. The support induction of \cite[Claims~3.5--3.6]{kps} establishes the two cases. A Find either gives support directly or supplies sufficiently light spokes. A valid Project lifts a child's witness or forces the next preference to be Find. Compute exactly which exceptional slots were hit. At a failed projection, return the record in its assigned round to its first invalid ancestor. The exact forest update there connects its endpoints at its weight. Earlier projections on its path were all valid, so this connection lifts to the root and persists under subsequent MSF updates. The final call keeps the root forest and unresolved marked identities. Each returned failure contributes to its ancestor's update and is omitted from this separate list of marked identities. This distinction prevents charging the same coarse mark for two inputs.

\paragraph{Counting exceptions.}
Normalize all counts by the edge count $e$ before the Prim contractions. Set
\[
 a=81L,\quad g=28+3\lceil\log_2L\rceil,\quad
 u=\rho/(\rho-1),\quad h=(1-\rho^2/(11/2))^{-1},\quad c_0=128/127.
\]
Let $E$ bound the number of coarse exceptions and $E_P$ the number occurring at a last Project. Let $F_P$ bound the latter count with each exception at depth $j$ weighted by $2(j+1)$. Write $B_0$ for a bound on the number of local exceptions. We obtain the bounds
\begin{align*}
 E&=445ha^2\sqrt N+\frac{c_0(J+1/J)u}{80L},&
 E_P&=445ha^2\sqrt N+\frac{c_0 u}{80LJ},\\
 F_P&=890h^2a^2\sqrt N+\frac{c_0u^2}{40LJ},&
 B_0&=438s\sqrt{aN}+2c_0/s.
\end{align*}
For a fixed coarse history on $x$ edges and $y$ vertices, the two respective bad-slot probabilities under independent round assignments are at most $(J+1/J)/t$ and $1/(Jt)$. The logarithm of the number of coarse histories is at most $(2+3\sqrt2)t^2\sqrt{xy}$, from the choices of spokes, forest and preferences. For the preferences, the vector of maxima of fixed subsets is determined by its histogram of winners. To see this, fix an ordering realizing one vector and sum the ranks of its winners. Any different vector of winners for the same subsets has a strictly smaller sum. Stars and bars bounds the number of such histograms. For the local histories, the analogous logarithm is at most $(2+\sqrt2)s\sqrt{xy}$ and there are at most two exceptional slots per edge. With exponent $\lambda=1/64$, the exponential moment for a fixed history and bad-slot probability $p_0$ is at most $\exp((\exp(\lambda)-1)xp_0)$. Adding these normalized moments over histories and both coarse event types gives a single assignment satisfying both bounds. Exhaustive search makes this choice deterministic. Using $(\exp(\lambda)-1)/\lambda<c_0$, summing square-root charges by Cauchy--Schwarz, and substituting $N/32^j$ gives the four displayed bounds. The factor $h$ sums the coarse square-root terms and $u$ sums the reciprocal block lengths. Ancillary proof note~02 supplies the history counts, the coarse and local sums, and the complete support induction.

\paragraph{Total size of the exact MSF calls.}
Put $R=\rho^4/32$, $S_i=\sum_{j\ge0}(j+1)^iR^j$ and
\[
 W=(64g/J^2+3)a^4NS_3,\qquad C=(64g/J^2)sa^4N .
\]
In a call on $y$ vertices, forest sizes and per-block declarations contribute at most $sW$ in aggregate. Before multiplying by the local factor and the bound for packet sizes, their count per call is at most $(2t^2+t+(t/J)k)y$. The quota bound $k\le(64/J)g(j+1)(a\rho^j)^3$ and the aggregate vertex bound give $W$ even with two additional factors of $j+1$ for packet sizes. The four contributions to $Y$ are the coarse exceptions, local exceptions, the root forest (at most $Ne$ edges), and the declarations and forests passed to intermediate calls. The return of failed projections has already been charged to the coarse exceptions. Hence
\[
                              Y/e\le E+B_0+N+sW.
\]

\subsection{Direct comparison costs}
It remains to bound $X$. We account separately for maintaining admissible records, selecting representatives of distinct labels, and identifying exceptions.

\paragraph{Maintaining admissible records.}
Store each owner list in fixed neighbor-label order. The bottom packets have size $b_0$. Form successively larger packets by joining four consecutive packets at a time, to tier $\ell=\lceil\log_2\max(1,j)\rceil$. A tier-$i$ packet maintains the lightest entry that passes its first $\min(j,2^i)$ projection predicates. Test the winner of its child tournament on the new strip, discarding and reporting a first-invalid entry when necessary. Each top champion is fully admissible. Building the tournaments consumes at most the one raw scan already paid. For a failure at predicate $i$ the strip tests and nonbottom repairs cost at most $5i$, apart from the $\log_2 b_0$ bottom repair. We charge an initial valid strip once per packet. A replacement's successful strips telescope along its path. Thus the normalized extra validation and repair charge is
\[
                 V_0=4/b_0+12saN+(\log_2 b_0)E_P+(5/2)F_P.
\]
The first terms cover full and partial packets, respectively. Ancillary notes~02 and~04 give the strip-by-strip charging argument.

\paragraph{Selecting distinct labels.}
Group the fully valid champions by label, keep the minimum of each group and build a binary minimum tournament on the remaining representatives. Up to $k$ extractions cost at most $k\lceil\log_2\max(1,h_0)\rceil$ on an owner list of length $h_0$. Reopen each mixed packet whose champion label is selected. Each label meets at most two mixed packets, one at each boundary of its fixed interval. An unopened packet with an unselected champion cannot contain an improving record, since $k$ distinct selected labels are at least as light. A pure packet with a selected champion cannot improve that representative. Keep the selected representatives, test all reopened candidates on their full chains, and repeat the label-grouped tournament on the survivors. This returns exactly the admissible $k$ lightest distinct labels. All new failures receive the return and support accounting above.

Top packets have at most $4b_0(j+1)^2$ entries. If $l_j$ is the number of owner lists at depth $j$ divided by $e$, then $l_j k\le C(j+1)R^j$. The two builds, extractions and explicit tests outside the initial raw scan cost at most
\begin{align*}
 O_0={}&1/b_0+sW+C\big[(2S_2-S_1)+8b_0(2S_4-S_3)\big]\\
      &+2C\big[(K+2)S_1+5(S_2-S_1)\big].
\end{align*}
Indeed the second list size is at most $(1+8b_0(j+1)^2)k$ and an item incurs at most $2j$ direct eligibility tests and one build comparison. For the two extraction tournaments, concavity bounds their cost at depth $j$ by $2kl_j[1+\log_2(1+1/l_j)]$ per original edge. Substitute $l_j\le s(a\rho^j/J)N/32^j$ by monotonicity. At that substituted value the bracket is at most $K+5j+2$. This yields the last term in $O_0$.

\paragraph{Identifying exceptional slots.}
Preferences depend only on the keys of the known forest edges and spokes, and on the thresholds in local batches. For $M_0$ prescribed pairs on $N_0$ keys, the deterministic local comparison lemma proved in ancillary note~03 uses at most $32768 M_0^{3/5}N_0^{2/5}$ comparisons. The proof enumerates compatible sign patterns on samples, finds common pivots for blocks of each candidate ordering, and resolves the remaining comparisons recursively. At density $M_0/N_0\ge2^{36}$, two stages leave one family on at most $.013N_0$ keys and two fallback lists with at most $M_0/8$ pairs each. Together with the constructive sample-cost bound, their normalized coefficients sum to at most one. Below that density direct comparisons suffice. For a coarse batch of $z$ queries, the directed forest path-maximum calculation~\cite{komlos} and the local comparison lemma run on at most $4z$ pairs of at most $2y$ keys. Check current and next-slot preferences on assigned queries and check the final slot on all edges in each block. The final-slot check identifies the last Find block and the next-slot check identifies the last Project. Local rounds need only compare endpoint thresholds. H\"older's inequality over time slices and the depth bound give the following normalized sweep charge:
\[
 H_0=\frac{N^{2/5}}{1-\rho/4}
       \left[131088\bigl(a/J+2\lceil a^{4/5}\rceil\bigr)
                   +32768s\lceil a^{2/5}\rceil\right]
                   +\frac{15a^2N}{1-\rho^2/32}.
\]
The term containing $a/J$ accounts for the whole-graph sweeps at block ends. Two families of $t^2$ groups yield the exponent $4/5$. The at most $t$ coarse batches containing local sweeps yield $2/5$. Ancillary notes~02--04 give the directed path requests and the summation.

Combining these charges yields
\[
                         X/e\le O_0+V_0+H_0+300sW+2aN .
\]
The last two terms pay for merging the existing representatives and updating spokes. The sums $S_i$ are rational functions of $R$. In particular, $S_4=(1+11R+11R^2+R^3)/(1-R)^5$. Substituting the two parameter choices and rounding each approximate root upward gives the bounds in Lemma~\ref{lem:finish}. The schedule depends only on the topology and comparison transcript. Every test of an undetermined order compares two original edge weights.\hfill$\square$

\clearpage
\section*{Historical comparison bounds}
Table~\ref{tab:history} gives explicit counts where available. An asymptotic running-time bound also bounds the number of weight comparisons, up to the constant implicit in the running time.

\begin{table}[!htb]
\centering
\caption{Edge-weight comparisons for MST and MSF. Upper bounds are deterministic and worst-case unless marked otherwise. The lower bounds concern worst-case construction on hard inputs.}\label{tab:history}
\small
\setlength{\tabcolsep}{4pt}
\renewcommand{\arraystretch}{1.14}
\begin{tabularx}{\linewidth}{@{}l>{\raggedright\arraybackslash}Xl@{}}
\toprule
Year & Algorithm or result & Weight comparisons\\
\midrule
1926 & Bor\r{u}vka, incidence scans~\cite{boruvka} & $\le (2m-n)\lfloor\log_2 n\rfloor$\\
1930/1957 & Jarn\'ik--Prim, array implementation~\cite{jarnik,prim} & $\le m-r+\binom{r}{2}$\\
1956 & Kruskal with mergesort~\cite{kruskal} & $\le mh-2^h+1$\\
1975/1976 & Yao and Cheriton--Tarjan~\cite{yao,ct} & $O(m\log\log n)$\\
1984/1987 & Fredman--Tarjan, Fibonacci heaps~\cite{ft} & $O(m\beta)$\\
1986 & Gabow--Galil--Spencer--Tarjan~\cite{ggst} & $O(m\max\{1,\log\beta\})$\\
1995 & Karger--Klein--Tarjan, randomized~\cite{kkt} & $O(m)$ in expectation\\
2000 & Chazelle, soft heaps~\cite{chazelle} & $O(m\alpha(m,n))$\\
2002 & Pettie--Ramachandran, optimal MSF~\cite{pr} & $O(T^*(m,n))$\\
2002 & Pettie--Ramachandran, MSF lower bound~\cite[Prop.~5.1]{pr} & $\ge m/2$\\
2012 & Prim with rank-relaxed weak heaps~\cite{weak} & $\le 3(m-r)+\tfrac32r\log_2 r$\\
2026 & Khanna--Putterman--Song~\cite{kps} & $O(m)$\\
2026 & Eriksson, explicit bound~\cite[Thm.~3(b)]{eriksson} & $\le 944m$\\
\midrule
2026 & \textbf{This work, upper bound on all graphs} & $\boldsymbol{<5.824m}$\\
2026 & \textbf{This work, lower bound on a sparse family} & $\boldsymbol{>1.641m}$\\
\bottomrule
\end{tabularx}
\par\smallskip
\begin{minipage}{\linewidth}\footnotesize
For the implementation-specific upper bounds, $n\ge2$, the graph is connected, $r=n-1$, and $h=\lceil\log_2 m\rceil$. Here $\alpha$ is an inverse Ackermann function, $\beta$ is the least $i\ge1$ with $\log_2^{(i)}n\le m/n$, and $T^*(m,n)$ is the optimal worst-case comparison depth for MSF over graphs with $m$ edges and $n$ vertices. The strict upper bound assumes $m>0$.
\end{minipage}
\end{table}

\paragraph{Derivation of the implementation-specific bounds.}
For Bor\r{u}vka's algorithm, scanning all incidences on a connected quotient with $e$ edges and $v$ vertices uses at most $2e-v$ comparisons. Each round reduces the number of vertices by at least half and removes at least one edge per eliminated vertex, so $2e-v$ never increases. For array-based Prim, the first edge reaching each of the $r$ other vertices initializes its key. At most $m-r$ later edges can require a comparison with a stored key, and selecting successive frontier minima costs at most $\binom r2$ comparisons. For Kruskal, balanced mergesort uses at most $mh-2^h+1$ comparisons, and testing for cycles uses only the endpoints. Finally, Prim with rank-relaxed weak heaps performs $r$ insertions, $r$ extractions, and at most $m-r$ decreases. The priority queue uses at most $2(m-r)+\tfrac32r\log_2 r$ comparisons~\cite[Theorem~4]{weak}. The $m-r$ relaxation tests give the bound in the table. For the early history of these algorithms, see Graham and Hell~\cite{history}.

\paragraph{Reproducibility.}
The accompanying source contains the exact parameter tables and three arithmetic checks. The lower-bound checker evaluates the hierarchy count with outward-rounded intervals. The upper-bound checkers verify the contraction schedules and the recursive comparison budget using rational arithmetic and rigorous enclosures. Instructions are in \texttt{anc/README.md}.

\paragraph{Acknowledgment.}
OpenAI's ChatGPT and Codex were used in developing the arguments, preparing the arithmetic certificates, and drafting the exposition.

\end{document}